\documentclass[conference]{IEEEtran}
\IEEEoverridecommandlockouts
\usepackage[T1]{fontenc}
\usepackage[utf8]{inputenc}
\usepackage{cite}
\usepackage{amsmath,amssymb,amsfonts}
\usepackage{graphicx}
\usepackage{textcomp}
\usepackage{xcolor}
\usepackage{booktabs}
\usepackage{array}
\usepackage{listings}
\usepackage{enumitem}
\usepackage{balance}
\usepackage{url}
\usepackage{hyperref}
\hypersetup{hidelinks}

\setlist{nosep,leftmargin=1.2em}

\definecolor{viol}{HTML}{C0504D}
\definecolor{sat}{HTML}{4F8A6B}
\definecolor{codebg}{HTML}{F5F5F2}

\newcommand{\viol}{\textcolor{viol}{\textbf{VIOLATION}}}
\newcommand{\nvo}{\textcolor{sat}{\textbf{NO\_VIOLATION\_OBSERVED}}}
\newcommand{\inc}{\textbf{INCONCLUSIVE}}
\newcommand{\B}{\textsf{B0}}
\newcommand{\M}{\textsf{M}}

\newcommand{\cve}[1]{\mbox{CVE-#1}}

\begin{document}

\title{Beyond the Patch: Reconstructing an Incomplete-Fix\\ Patch Bypass in LLM Agent Frameworks with\\ Control-Gated, Benign Probes}

\author{
\IEEEauthorblockN{Chaitanya Garware}
\IEEEauthorblockA{%
\textit{Independent Researcher}\\
ORCID: 0009-0004-8770-8825\\
\texttt{chaitanyagarware7@gmail.com}}
}

\maketitle
\begin{abstract}
\noindent
When a vulnerability at a security boundary of an LLM agent framework is patched,
the advisory identifies the affected and patched versions, and replaying the
proof-of-concept (PoC) tests only whether its \emph{reported trigger} is still
reproducible. That the trigger no longer fires does not, on its own, show that the
underlying \emph{security invariant} holds on every code path that reaches the
boundary. We present a small, control-gated
reconstruction method and a pilot study built around one question: can
original-trigger replay report a boundary as fixed while the same invariant stays
violable through a distinct, separately reachable path at the same version? We
compare original-trigger replay (\B) with a distinct sibling-path probe (\M) for
PraisonAI, and perform five additional before/after reconstructions across
Langroid, Agno, and CrewAI. Each run emits a tri-state verdict of \viol, \nvo, or
\inc. A mandatory positive (authorized) control gates the verdict,
so that a non-violation result is admitted only when the control passes in the same
run. All probes are benign, using an in-memory flag, a filesystem sentinel, or a
builtins-reachability check, and none performs a real payload. Our central result
reproduces an \emph{officially recognized} incomplete fix. In PraisonAI,
original-trigger replay reports the tool-loading boundary as fixed at
version~4.6.31, yet the invariant is still violated through the
\texttt{tool\_override} path. That residual was later assigned its own identifier
(\cve{2026-44334}, catalogued as a ``\cve{2026-40287} patch bypass'') and closed at
4.6.32. The five before/after measurements span four frameworks and four invariant
classes. For a CVE whose advisory records no fixed version, we observe confinement
in tested release 1.11.1 after observing a violation in 1.11.0. We are explicit
about scope. A
non-violation verdict means ``no violation observed under the tested probe,
version, and configuration,'' not a proof of safety. The headline harness is
control-gated per path, and fourteen fault-injection scenarios produced their
expected verdicts. Only one same-version \B$\neq$\M{} divergence is demonstrated.
The PraisonAI probe invokes the documented loader functions directly, which
establishes boundary behavior under the supplied setup rather than a full
end-to-end exploit. Finally, we do \emph{not} claim the method is a new testing
technique or that it estimates prevalence.
\end{abstract}

\section{Introduction}
LLM \emph{agent frameworks} route model- and attacker-influenced input into
classic software boundaries: file access, expression evaluation, and tool
dispatch. A PRISMA review of agentic-LLM security finds that action-layer
vulnerabilities (tool misuse, code injection, sandbox escape) appear in only
$4.7\%$ of papers despite their real-world weight~\cite{agenticsok}. Broader
surveys note that tool bridges ``expose classic bugs (code exec, traversal,
SSRF) when schemas/validation are weak''~\cite{agentsok2}.

When such a bug is fixed, an advisory records a \emph{fixed} version and a PoC
documents the \emph{reported trigger}. The empirical patch literature shows this
is an incomplete form of assurance: an initial patch ``may fail to fully
eliminate the vulnerability or may introduce new vulnerabilities or bugs that
require subsequent fixes''~\cite{multipatch}, and a large kernel study catalogs
thousands of incomplete fixes that each required follow-ups~\cite{ispkernel}.
The same work prescribes what good fix-testing should check: that ``the PoC can
no longer be triggered \emph{and} that normal functionality is
preserved''~\cite{multipatch}.

Agent-framework boundaries are a sharp setting because they are reached through
\emph{several} code paths: a local tool loader and a template-override loader; a
default-sanitized evaluator and a documented trusted-mode evaluator; a read tool
and a write tool. A patch that neutralizes the reported path can leave a sibling
path open while the advisory still reads ``fixed.'' PraisonAI is a documented
example. The fix for \cve{2026-40287} added an opt-in gate to two tool-loading
sinks, but an equivalent sink in \texttt{templates/tool\_override.py} was missed
and later assigned \cve{2026-44334}, catalogued as a ``\cve{2026-40287} patch
bypass''~\cite{cve44334}.

\paragraph{This paper.}
We build a deliberately small reconstruction method around that phenomenon and
apply it to a hand-selected pilot corpus. We compare original-trigger replay (\B)
with a distinct sibling-path probe (\M) for PraisonAI, and perform five additional
before/after reconstructions across Langroid, Agno, and CrewAI, each emitting a
tri-state verdict gated by a mandatory positive control.
Our aim is not a new testing technique; positive controls, differential testing,
and variant analysis are all established. The contribution is instead (i)~a
repeatable, benign, control-gated \emph{reconstruction} that cleanly separates
``reported trigger blocked'' from ``invariant holds,'' and (ii)~an empirical
demonstration and set of measurements on real agent-framework CVEs.

\paragraph{Contributions.}
\begin{itemize}
\item \textbf{A control-gated, tri-state reconstruction} that withholds a
non-violation verdict unless a positive control passes (every broken condition we
tested yields \inc), with benign effect observers (\S\ref{sec:method}).
\item \textbf{A reproduction of an officially recognized incomplete fix} as a
same-version \B$\neq$\M{} divergence in PraisonAI, with the positive control
doubling as reachability evidence (\S\ref{sec:praison}).
\item \textbf{Five before/after measurements} across Langroid, Agno, and CrewAI,
each a benign, control-gated reconstruction of a disclosed CVE, with the eval and
dispatch cases verified end-to-end through the real input-parsing pipeline
(\S\ref{sec:measurements}). We also report \textbf{a fix-provenance result}: for a
CVE whose advisory records no fixed version, we observe confinement in tested
release 1.11.1 after observing a violation in 1.11.0 (\S\ref{sec:prov}).
\item \textbf{A benign artifact and reproducibility manifest} (\S\ref{sec:avail}).
\end{itemize}
We frame this as a \emph{method-plus-case-study pilot}, not a population study,
and we keep our non-claims explicit throughout.

\section{Background and Threat Model}
\label{sec:background}
\paragraph{Boundaries and invariants.}
We study three boundaries. For \emph{tool loading}, a named tool is resolved to
code, sometimes from user-reachable locations; the invariant is that untrusted
tool references must not load or execute local code unless an operator has opted
in. For \emph{expression evaluation}, a model-produced expression is evaluated;
the invariant, in the configuration the advisory targets, is that evaluation must
not reach dangerous builtins. For \emph{file tools and dispatch}, the invariants
are read/write confinement to a workspace and dispatch only from an authorized
sender. We treat each invariant as a property that should hold \emph{independent of
the path} reaching the boundary.

\paragraph{Attacker model.}
The attacker influences the inputs that reach a boundary, namely tool
names and arguments, an evaluated expression, a file path, or the apparent sender.
This is consistent with indirect prompt injection and untrusted-content
settings~\cite{protocol,coding}. The attacker cannot modify framework source. We
measure whether the framework \emph{enforces} the invariant when such an input
arrives; we do not measure how often a model would emit one.

\paragraph{Test setup vs.\ attacker capability.}
We distinguish researcher-controlled setup from attacker capability. For the
PraisonAI boundary, the planted \texttt{tools.py} in the working directory models
the attacker-influenced content that the \emph{documented} loader auto-imports;
placing it is part of our test harness, not a claim that the attacker writes to
the host. Our probe invokes the documented loader functions
(\texttt{ToolResolver.\_load\_local\_tools}, \texttt{create\_tool\_registry\_with\_overrides})
\emph{directly}. This establishes the boundary's behavior under the supplied
setup; it does not reproduce the full HTTP-request-to-server exploit chain the
advisory describes, and we do not claim end-to-end exploitation for this case
(the Langroid eval/dispatch cases are additionally driven through the real
input-parsing pipeline; \S\ref{sec:measurements}). Where an invariant is
\emph{configuration-dependent} (for example, a documented trusted mode), we report
the configuration the relevant advisory names and do not transfer a verdict across
configurations (\S\ref{sec:threats}).

\section{Method}
\label{sec:method}
\begin{figure}[t]
\centering
\includegraphics[width=\linewidth]{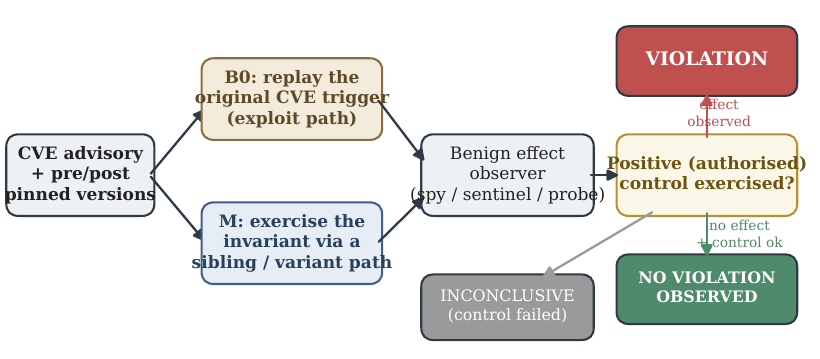}
\caption{The reconstruction replays the original trigger (\B) and exercises the
invariant through a second in-repo path (\M). A benign observer reports whether
the invariant was violated; a non-violation verdict is admitted only if a
positive (authorized) control succeeds in the same run, otherwise the run is
\inc.}
\label{fig:overview}
\end{figure}

\paragraph{Objects and verdict.}
For a boundary $b$ at version $v$, a \emph{probe} is a benign action that
succeeds \emph{iff} the invariant is violated (reading a file planted outside the
workspace; resolving \texttt{\_\_import\_\_} inside an evaluated expression). A
probe is paired with a \emph{positive control}: an authorized action at the same
boundary that must succeed if the boundary is wired correctly. The run reports:
\[
\resizebox{0.98\columnwidth}{!}{$\displaystyle
\text{verdict}(b,v)=
\begin{cases}
\inc & \text{control fails}\\
\viol & \text{control ok}\wedge\text{effect seen}\\
\nvo & \text{control ok}\wedge\text{no effect}
\end{cases}$}
\]

\noindent\fbox{\parbox{0.93\linewidth}{\small
\textbf{Meaning of \nvo.} The probe exercised the boundary (the positive control
succeeded) and produced no violating effect \emph{under this probe, version, and
configuration}. It is \emph{not} a proof that the boundary is free of all
bypasses. We use this reading everywhere, and we avoid ``complete fix,''
``certifies,'' and ``secure.''}}

\medskip
The control gate is the methodological core: a boundary that threw on
\emph{all} input, including the authorized control, would otherwise masquerade as
fixed. The verdict also deliberately maps an infrastructure failure to \inc{}
rather than to a security outcome. We discuss the tradeoff in \S\ref{sec:threats},
since a directly observed unauthorized effect with a failing unrelated control is
reported as \inc, which is conservative and can suppress a true violation.

\paragraph{Per-path gating and harness validation.}
Each path is gated by its \emph{own} positive control: a \nvo{} on \B{} requires
\B's control to pass, and likewise for \M. For the PraisonAI harness the probe
writes two markers, a \emph{reached} sentinel (proof the loader executed the
planted code) and an \emph{effect} marker. A sink that runs but whose effect
is not observed, for example when the planted code raises or the observer cannot
record, is reported \inc{} rather than non-violation. Measurements run one per
process to avoid module-cache and environment-flag leakage, and the verdict logic
is a \emph{single shared module} used by both the live harness and a
fault-injection suite. That suite exercises the actual harness under fourteen
scenarios. Two are \emph{healthy} (vulnerable\,$\to$\,\viol,
patched\,$\to$\,\nvo); the remaining twelve are faults: each path's control
failing, a missing entry point, an unrelated probe exception, an observer failure,
both stale-marker cases (a blocked path with a leftover marker, and a vulnerable
path that overwrites it), an empty record, a mistyped boolean, a
missing/erroneous version, a wrong expected version, and a subprocess failure.
Observation is token-based: a fresh per-run token must match the marker's content,
so a leftover marker is never counted as an effect. All fourteen scenarios
produced their expected verdicts. Every genuinely broken condition yields \inc{}
(never a non-violation), the blocked stale case yields \nvo{}, and the overwrite
case yields \viol{} (\S\ref{sec:praison}).

\paragraph{\B{} vs.\ \M.}
\B{} re-expresses the public PoC: the exact path the CVE used. \M{} reaches the
same invariant through an independent, in-repo path, a sibling loader or a second
tool action, chosen by reading source rather than by mutating the PoC string.
We distinguish a \emph{distinct code path} (a different function or sink, ideally
one with its own provenance such as a separate CVE) from a mere \emph{input
variant} of the same path, and we label each case accordingly
(\S\ref{sec:measurements}).

\paragraph{Benign effect observers.}
Three observers suffice: an \emph{in-memory flag} recording whether an inert
handler was invoked; a \emph{filesystem sentinel} (a uniquely named marker
outside the intended directory); and a \emph{builtins-reachability} check that
resolves \texttt{\_\_import\_\_} \emph{without calling it}. Probes plant inert
marker modules or inert tool objects. Loading such a module does execute benign
Python, but no probe performs a network connection, spawns a subprocess, or runs
a payload (\S\ref{sec:ethics} states what we did and did not monitor).

\paragraph{Reconstruction protocol.}
For each boundary we (1)~read the advisory and locate the fixing change in
source; (2)~pin the last-affected and first-fixed versions, verifying the pin
against the package index \emph{and} the version-control history; (3)~install
each version in an isolated interpreter and run the probe and its positive control
(for PraisonAI, the distinct \B{} and \M{} paths; for the five before/after cases,
an input-variant probe at each version), printing the resolved package version the
process actually imported; and (4)~record the verdict and the raw output.

\section{Corpus and Setup}
\label{sec:corpus}
\begin{table*}[t]
\centering\small
\caption{Pilot corpus. One same-version \B$\neq$\M{} divergence (PraisonAI) plus
five before/after reconstructions, across four frameworks and four invariant
classes. $^{\dagger}$Author's own reported advisory, used only as a calibration
target. ``\M{} type'' distinguishes a genuinely distinct code path from an input
variant of the reported path.}
\label{tab:corpus}
\setlength{\tabcolsep}{4pt}\renewcommand{\arraystretch}{1.15}
\begin{tabular}{@{}l l p{2.35cm} p{1.75cm} p{2.05cm} p{2.3cm} p{1.2cm}@{}}
\toprule
\textbf{ID} & \textbf{Framework} & \textbf{Component} & \textbf{Invariant} &
\textbf{Pre\,$\rightarrow$\,Post} & \textbf{CVE(s)} & \textbf{\M{} type} \\
\midrule
B1 & PraisonAI & tool loading & load gate & 4.5.139, 4.6.31 $\rightarrow$ 4.6.32 & \cve{2026-40287} / \cve{2026-44334} & distinct path \\
B2$^{\dagger}$ & Langroid & file tools & path confinement & 0.63.0 $\rightarrow$ 0.64.0 & \cve{2026-50181} & input variant \\
B3 & Langroid & \texttt{pandas\_eval} & eval-globals (\texttt{full\_eval}) & 0.65.1 $\rightarrow$ 0.65.2 & \cve{2026-54769} & input variant \\
B4 & Langroid & dispatch & sender verification & 0.65.2 $\rightarrow$ 0.65.3 & \cve{2026-54771} & input variant \\
B5 & Agno & \texttt{PythonTools} & path confinement & 2.3.23 $\rightarrow$ 2.3.24 & \cve{2026-76832} & input variant \\
B6 & CrewAI & \texttt{FileWriterTool} & path confinement & 1.11.0 $\rightarrow$ 1.11.1 & \cve{2026-37007} & input variant \\
\bottomrule
\end{tabular}
\end{table*}

Table~\ref{tab:corpus} lists the corpus. This is a \emph{purposive} sample, not
a random one. Frameworks were chosen for a 2025--2026 CVE at one of our three
boundaries \emph{and} a reconstructible, offline-testable sink with a pinnable
pre/post pair. Only PraisonAI (B1) provides a distinct second code path with its
own CVE; the other five are before/after reconstructions in which \M{} probes
beyond the literal PoC string (absolute paths, a second action) on the same
path. We do not infer prevalence or detection accuracy from this sample
(\S\ref{sec:threats}).

\paragraph{Environment.}
Each version runs in its own virtual environment, and each test prints the
resolved package version and imported module path. The six-boundary corpus was run
under CPython~3.12 on Windows. The corrected PraisonAI matrix and its
fault-injection suite were additionally executed under CPython~3.13 on Linux
(Ubuntu~24.04), reproducing the same verdicts, with \texttt{praisonaiagents}
pinned at 1.7.23. The accompanying manifest (\S\ref{sec:avail}) records
interpreter, OS, \texttt{pip freeze}, module paths, and commands.

\section{Results}
\label{sec:results}
\begin{figure*}[t]
\centering
\includegraphics[width=0.92\textwidth]{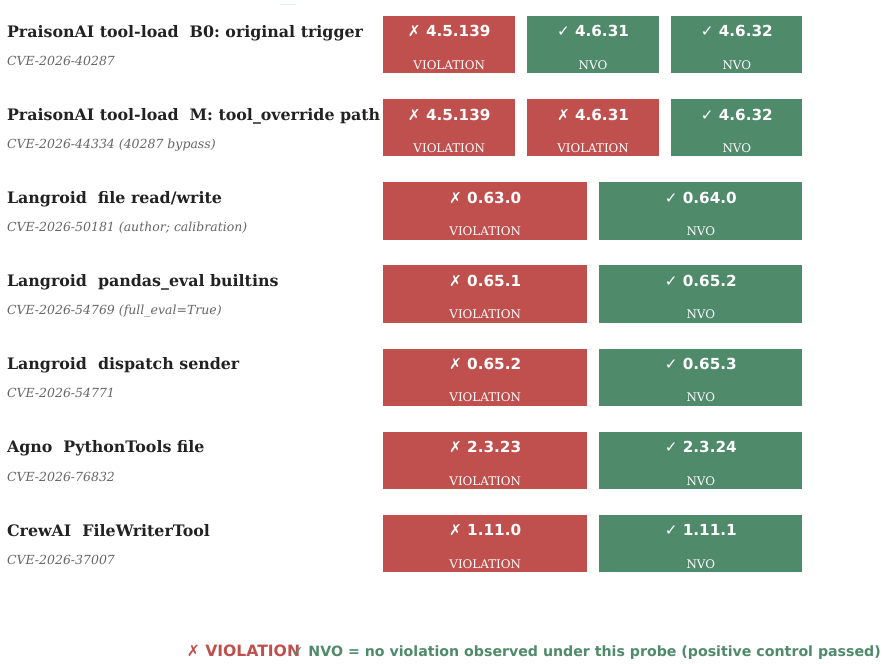}
\caption{Per-boundary verdicts. Each row carries its own version axis. The two
PraisonAI rows show the same boundary tested by \B{} (original trigger) and
\M{} (\texttt{tool\_override}); they diverge at 4.6.31. A green cell denotes
\nvo{} (no violation observed under the tested probe; positive control passed),
not a proof of safety.}
\label{fig:results}
\end{figure*}

Figure~\ref{fig:results} summarizes every verdict.

\subsection{Central result: a reproduced patch bypass (PraisonAI)}
\label{sec:praison}
PraisonAI resolves tools through more than one path. \B{} replays the
\cve{2026-40287} trigger (the \texttt{tool\_resolver} local-tool path), and \M{}
exercises a sibling path, \texttt{create\_tool\_registry\_with\_overrides}, which
loads a tool via a \texttt{tool\_override}. The two paths are guarded
independently. This is not our interpretation. The advisory for \cve{2026-44334}
is catalogued as a ``\cve{2026-40287} patch bypass'' and states that the
\cve{2026-40287} fix gated the local-tool import in \texttt{tool\_resolver.py}
and \texttt{api/call.py}, while the equivalent sink in
\texttt{templates/tool\_override.py} was not patched; it was fixed in
4.6.32~\cite{cve44334}.

\begin{figure}[t]
\centering
\includegraphics[width=\linewidth]{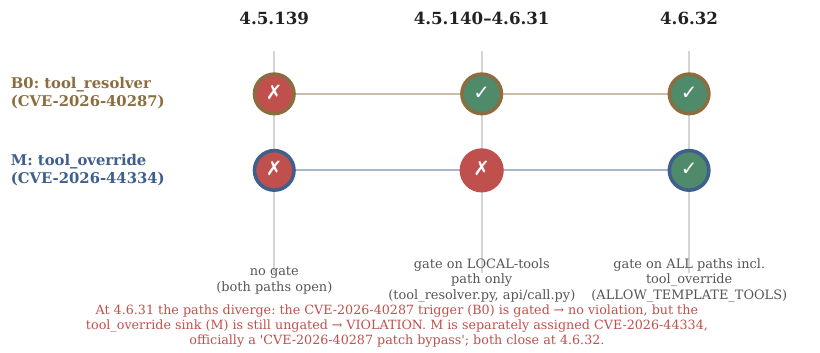}
\caption{PraisonAI tool-loading verdicts. \B{} (\cve{2026-40287}) becomes \nvo{}
once its path is gated, while \M{} (\texttt{tool\_override}, separately assigned
\cve{2026-44334}) remains \viol{} until 4.6.32.}
\label{fig:praison}
\end{figure}

Our corrected harness gates each path with its own positive control and runs one
measurement per process (\S\ref{sec:method}). With the local-tool opt-in
\emph{off}: at 4.5.139 both paths execute the planted inert module (\B{} and \M{}
\viol); at 4.6.31, \B{} returns \nvo{} because the \texttt{tool\_resolver} path is
gated, while \M{} returns \viol{} because the \texttt{tool\_override} sink still
executes; at 4.6.32 both return \nvo{} (Fig.~\ref{fig:praison}). The per-path
control doubles as reachability evidence: with opt-in \emph{on}, each path executes
the module and writes both markers, which confirms the probe reaches the real
sink. These verdicts were produced by the gated harness on Linux/CPython~3.13 and
match the earlier Windows/3.12 run. Against this same harness, all fourteen
fault-injection scenarios produced their expected verdicts (\S\ref{sec:method}):
every genuinely broken condition returns \inc{}, and the two stale-marker cases
resolve correctly (blocked\,$\to$\,\nvo, overwrite\,$\to$\,\viol), so the recorded
\B{}/\M{} verdicts are not artifacts of a silently broken test. As noted in
\S\ref{sec:background}, the probe invokes the documented loaders directly, so it
establishes the boundary behavior the advisory describes, not the full HTTP
exploit chain. A PoC-replay view would report the boundary fixed at 4.6.31; the
invariant-guided view does not, and the separate assignment of \cve{2026-44334}
corroborates that the residual was a real, recognized vulnerability.

\paragraph{Gate provenance (verified).} Pulling \texttt{tool\_resolver.py}
directly from the distributed PyPI wheels shows the \cve{2026-40287} opt-in guard
\emph{absent} at 4.5.139 and \emph{present} at 4.5.140 and 4.6.31 (0, 3, and 3
textual occurrences of \texttt{PRAISONAI\_ALLOW\_LOCAL\_TOOLS}; at 4.5.139
\texttt{\_load\_local\_tools} calls \texttt{exec\_module} with no guard). In the
shipped artifacts the \B{} gate therefore first appears at \textbf{4.5.140},
consistent with our executed baseline (\B{} \viol{} at 4.5.139). We note a
metadata/artifact discrepancy: the advisory lineage labels 4.5.139 as the
\cve{2026-40287} fix release, one patch version earlier than where the guard
appears in the wheel. The \M{} residual (\cve{2026-44334}) is ungated across this
whole span; the advisory's affected range is $\ge$4.5.139, $<$4.6.32 (fixed
4.6.32), which our 4.6.31~\M{}~\viol{} and 4.6.32~\M{}~\nvo{} measurements match.

\subsection{Before/after reconstructions}
\label{sec:measurements}
Five boundaries are version before/after pairs. Each positive control passed, so
each \nvo{} reflects an exercised boundary.
\begin{itemize}
\item \textbf{Langroid \texttt{pandas\_eval} (B3, \cve{2026-54769}).} The advisory
describes a sandbox escape to RCE via an ``incomplete \texttt{eval()}
mitigation,'' exploitable when \texttt{full\_eval=True}; our probe resolves
\texttt{\_\_import\_\_} under that documented configuration. \viol{} at 0.65.1;
\nvo{} at 0.65.2, where \texttt{safe\_eval\_globals} walls off builtins.
\item \textbf{Langroid file tools (B2, \cve{2026-50181}).} Traversal read/write at
0.63.0 (\viol); confined at 0.64.0 (\nvo). This is the author's own reported CVE
and is used only for calibration.
\item \textbf{Langroid dispatch (B4, \cve{2026-54771}).} An inert tool dispatched
with a forged \texttt{USER} sender fires at 0.65.2 (\viol) and is rejected at
0.65.3 (\nvo); a legitimate \texttt{LLM}-sender dispatch (positive control) fires
in both.
\item \textbf{Agno \texttt{PythonTools} (B5, \cve{2026-76832}).} Traversal read
escapes \texttt{base\_dir} at 2.3.23 (\viol); confined at 2.3.24 (\nvo).
\item \textbf{CrewAI \texttt{FileWriterTool} (B6, \cve{2026-37007}).} Traversal
write escapes the target directory at 1.11.0 (\viol); rejected at 1.11.1 (\nvo).
\end{itemize}
For B3 and B4 we also drive the \emph{full} input-parsing pipeline. An
attacker-shaped \texttt{ChatDocument} whose \texttt{content} is the tool JSON is
passed to the agent's normal \texttt{handle\_message}, which parses the tool from
content and dispatches it, with no object pre-injected. The end-to-end verdicts
match the boundary-level ones (B3: \viol{} at 0.65.1, \nvo{} at 0.65.2; B4:
\viol{} at 0.65.2, \nvo{} at 0.65.3), with the positive control (a benign
expression; a legitimate \texttt{LLM}-origin dispatch) passing throughout. This
establishes that an attacker-controlled input reaches the patched sink, not only
that the sink behaves correctly when called directly.

\subsection{Fix-provenance for a CVE with no recorded fixed version (CrewAI)}
\label{sec:prov}
For \cve{2026-37007} (MITRE-assigned), the CVE record references only a fix commit
(\texttt{crewAIInc/crewAI} \texttt{713fa7d}) and gives no fixed package
version~\cite{cve37007}. Third-party databases infer the affected range as
``$\le$1.10.2rc1'' (retrieved 2026-11-15), that is, fixed in any later release. We
find that inference incorrect. At 1.11.0, a stable release after 1.10.2rc1,
\texttt{FileWriterTool.\_run} still joins the attacker-controlled filename
directly, and our run is \viol:
\begin{lstlisting}
# crewai-tools 1.11.0 (still vulnerable)
filepath = os.path.join(kwargs.get("directory") or "",
                        kwargs["filename"])
with open(filepath, mode) as f: ...
\end{lstlisting}
The confinement check first appears in 1.11.1, where our run is \nvo:
\begin{lstlisting}
# crewai-tools 1.11.1 (fixed)
real_fp = Path(filepath).resolve()
if not real_fp.is_relative_to(Path(directory).resolve()):
    return "Error: ... must not escape the target directory"
\end{lstlisting}
We observe confinement in tested release 1.11.1 (\nvo) after observing a violation
in 1.11.0 (\viol). We do not claim 1.11.1 is the \emph{earliest} fixed release,
since we did not test every intervening prerelease. We make \emph{no} claim of an
official metadata error. The official advisory asserts no fixed version, and the
inaccurate ``after 1.10.2rc1'' guidance is a third-party aggregator's. We plan to
notify the maintainers and the aggregator; at time of writing no report has been
sent and none is acknowledged.

\paragraph{LlamaIndex (excluded).} During screening we examined a LlamaIndex
advisory (\texttt{encode\_image} path traversal, recorded fixed at
\texttt{llama-index-core}~0.12.41) and found the cited function byte-identical
between 0.12.40 and 0.12.41. A byte-identical function does not prove an
ineffective patch, since the change could be in a caller, helper, or packaging.
We therefore treat this case as \emph{unresolved} and exclude it from the corpus,
and we make no claim about it.

\section{Discussion}
\label{sec:discussion}
\paragraph{What the \B$\neq$\M{} split buys.}
PoC replay answers ``is the reported trigger blocked?''; it does not answer
``does the boundary enforce its invariant on other paths?''. The PraisonAI case
shows the gap is real and consequential, and the independent assignment of
\cve{2026-44334} shows it is recognized by the ecosystem, not merely by us. The
additional work is reading the fixing diff to find a second path and writing one
benign probe plus one control per path. We did not measure analyst time or
runtime, so we make no quantitative cost claim.

\paragraph{A configuration-dependent observation, separated.}
We also ran the \texttt{full\_eval=True} builtins probe on earlier Langroid
releases (0.59.31, 0.59.32) and observed violations. We do \emph{not} count these
as incomplete fixes. The earlier advisories (\cve{2025-46724}, \cve{2026-25481})
hardened the \emph{default-mode} sanitizer, a different configuration, and
trusted-mode permissiveness there was not the behavior those fixes targeted.
\cve{2026-54769} is the advisory that reclassified trusted-mode builtins exposure
as a vulnerability, and B3 reconstructs exactly that. We mention the earlier
observations only as configuration-dependent context.

\section{Limitations and Threats to Validity}
\label{sec:threats}
\textbf{Meaning of non-violation.} \nvo{} means ``no violation observed under the
tested probe, version, and configuration,'' never a proof of safety; a stronger
probe could reveal a bypass a weaker one missed.

\textbf{One same-version divergence.} The \B$\neq$\M{} claim is demonstrated for a
single boundary (PraisonAI). The other five are before/after reconstructions, not
sibling-path divergences, and we label them as input-variant probes
(Table~\ref{tab:corpus}). The central claim is therefore a case study with
supporting measurements, not a generalization.

\textbf{Direct invocation (PraisonAI).} The PraisonAI probe calls the documented
loader functions directly and plants \texttt{tools.py} as part of the test setup.
It establishes boundary behavior under that setup, not the full
HTTP-request-to-server exploitation the advisory describes. The Langroid eval and
dispatch cases are additionally driven through the real input-parsing pipeline,
while the file-confinement cases (B2, B5, B6) are direct tool-API calls.

\textbf{Harness guarantees are bounded.} The control gate and the fourteen-scenario
fault suite show the harness returns \inc{} on every broken condition we tested.
This is evidence, not a proof that \emph{no} broken condition can yield a false
non-violation.

\textbf{Sink reachability for B3/B4 (validated).} We drove the real parse/dispatch
pipeline with an attacker-shaped \texttt{ChatDocument} whose \texttt{content} is
the tool JSON, fed to the normal \texttt{handle\_message}, and obtained the same
verdicts as the boundary-level probes (\S\ref{sec:measurements}). This establishes
end-to-end reachability, and the injection-based probes are retained as faster
equivalents.

\textbf{Verdict conservatism.} Mapping a failing positive control to \inc{} can
suppress a directly observed violation when an unrelated dependency fails. We
accept this conservative bias to avoid false non-violations.

\textbf{Configuration scope.} B3 is measured under \texttt{full\_eval=True}, the
configuration its advisory names, and we do not transfer the verdict to default
mode.

\textbf{Harness validation.} The primary check is the fourteen-scenario
fault-injection suite run against the actual PraisonAI harness
(\S\ref{sec:method},~\S\ref{sec:praison}): every genuinely broken condition yields
\inc{} and both stale-marker cases resolve correctly. As an independent
corroboration on a second framework, we also ran the real Agno boundary under four
deliberately broken conditions (wrong package import, forced control failure, an
unreachable sink method, and a fixed-name stale marker). The first three each
yielded \inc{}, and the stale-marker case exposed the false \viol{} that a
non-unique sentinel would cause, the defect our per-run unique token now avoids.

\textbf{Platform.} Path-confinement cases (B2, B5, B6) were run on Windows only,
and symlink and path semantics differ on other platforms. A Linux replication is
left to future work, as our host's WSL Python version was incompatible with the
pinned framework releases.

\textbf{Conflict of interest.} The author reported \cve{2026-50181} (B2). It is
used only for calibration, and no central claim depends on it.

\textbf{Reproduction and environments.} All runs are the author's own
(AI-assisted). The six-boundary corpus was executed on a Windows host under
CPython~3.12, and the corrected PraisonAI matrix and its fourteen-scenario fault
suite were executed in a Linux container (Ubuntu~24.04, CPython~3.13), with
identical verdicts. These are author-run reproductions on two of the author's own
environments. No independent third-party reproduction has been performed, and we
do not describe author/AI-assisted runs as such.

\section{Related Work}
\label{sec:related}
Empirical studies establish that security fixes are frequently incomplete and
prescribe richer fix-testing~\cite{multipatch,ispkernel}. We operationalize that
prescription for agent-framework boundaries with a control-gated, benign probe
and a \B{}-vs-\M{} split, and we reproduce one ecosystem-recognized instance.
Systematizations map the agentic attack surface and note the under-study of the
action layer~\cite{agenticsok,agentsok2}; taxonomies of defense failures name
incomplete isolation as a recurring flaw~\cite{ipitax}; and threat models for tool
and protocol layers motivate our attacker model~\cite{protocol,coding}. Our
contribution is orthogonal to new attacks or defenses: it is a method to check,
and a measurement of whether, a \emph{fix} restored a boundary's invariant.
Relative to standard security regression testing, the specific elements we rely
on are the mandatory positive control (which also serves as reachability
evidence) and the explicit distinct-path-versus-input-variant labeling, and we do
not claim these as novel techniques.

\section{Ethics and Disclosure}
\label{sec:ethics}
All studied vulnerabilities are public and fixed in a later release. Probes are
benign: they plant inert marker modules or inert tool objects and observe an
in-memory flag, a filesystem sentinel, or builtins reachability. Loading an inert
module executes benign Python. We did not run under OS-level network/process
monitoring, so our ``no network/no subprocess'' statement describes what the
probes are written to do, not a monitored measurement. A further, \emph{unfixed}
PraisonAI sibling sink (in \texttt{agents\_generator.py}) is documented publicly.
We did not build a probe for it, as it is unpatched, and any examination would
follow coordinated disclosure in an isolated environment. We plan to notify the
CrewAI maintainers and aggregator of the fix-version precision; no disclosure has
been sent or acknowledged at submission.

\section{Availability}
\label{sec:avail}
The per-boundary tests, a \texttt{results} log, and a reproducibility manifest
(interpreter and OS versions, \texttt{pip freeze} per environment, exact
commands, resolved package versions, and raw stdout) are maintained separately as
a benign artifact and are available from the author on request. We do not claim a
public release or independent reproduction; for a venue requiring it, the artifact
can be anonymized and deposited.

\section{Conclusion}
Advisories identify affected and patched versions, and replaying a PoC shows
whether its trigger is still reproducible. Neither establishes that a boundary's
invariant is restored on every path. A small, benign, control-gated reconstruction
that replays the original trigger and independently exercises a second path
reproduces an officially recognized incomplete fix in PraisonAI, and it supports
five before/after measurements and one fix-provenance result across four
frameworks. The method is small and the corpus is a pilot. Its value is a
discriminating check that a fix restored its invariant, and evidence that this
check catches residuals that PoC replay and advisory labels miss.

\balance


\begin{thebibliography}{99}
\small
\bibitem{multipatch} Why Not Fix It Once and for All? An Empirical Study of
Multiple Patches for Vulnerability Fixes in Open-Source Software.
\emph{arXiv:2607.13206}, 2026.
\bibitem{ispkernel} Characteristics, Root Causes, and Detection of Incomplete
Security Bug Fixes in the Linux Kernel. \emph{arXiv:2511.17799}, 2025.
\bibitem{agenticsok} On Understanding, Identifying, and Mitigating
Vulnerabilities in Agentic Large Language Models. \emph{arXiv:2608.10530}, 2026.
\bibitem{agentsok2} Systematization of Agentic-LLM Security. \emph{arXiv:2603.22928}, 2026.
\bibitem{ipitax} Z.~Ji et al. Taxonomy, Evaluation and Exploitation of
IPI-Centric LLM Agent Defense Frameworks. \emph{arXiv:2511.15203}, 2025.
\bibitem{protocol} M.~A.~Ferrag et al. From Prompt Injections to Protocol
Exploits. \emph{arXiv:2506.23260}, 2025.
\bibitem{coding} Prompt Injection Attacks on Agentic Coding Assistants.
\emph{arXiv:2601.17548}, 2026.
\bibitem{cve40287} \cve{2026-40287} / GHSA-g985-wjh9-qxxc, PraisonAI local-tool
auto-import gate.
\bibitem{cve44334} \cve{2026-44334} / GHSA-xcmw-grxf-wjhj, ``PraisonAI has
unauthenticated RCE via \texttt{tool\_override.py} (\cve{2026-40287} patch
bypass),'' affected $\ge$4.5.139 $<$4.6.32, fixed 4.6.32. Retrieved 2026-11-15.
\bibitem{cve50181} \cve{2026-50181} / GHSA-fg23-3346-88f5, Langroid file-tool
path traversal (fixed 0.64.0).
\bibitem{cve54769} \cve{2026-54769} / GHSA-q9p7-wqxg-mrhc, ``Langroid: Sandbox
Escape to RCE via Incomplete \texttt{eval()} Mitigation,'' \texttt{full\_eval=True},
affected $<$0.65.2, fixed 0.65.2. Retrieved 2026-11-15.
\bibitem{cve54771} \cve{2026-54771}, Langroid tool-dispatch sender verification
(fixed 0.65.3).
\bibitem{cve76832} \cve{2026-76832}, Agno \texttt{PythonTools} path traversal
(fixed 2.3.24).
\bibitem{cve37007} \cve{2026-37007} (MITRE-assigned, CWE-22), CrewAI
\texttt{FileWriterTool} path traversal; CVE record references fix commit
\texttt{crewAIInc/crewAI 713fa7d} and no fixed package version. Fix first observed
by this work in \texttt{crewai-tools} 1.11.1 (1.11.0 still vulnerable). Retrieved
2026-11-15.
\end{thebibliography}
\end{document}